\documentclass[twocolumn,aps,prl,superscriptaddress]{revtex4-2}
\usepackage[T1]{fontenc}
\usepackage{babel}
\usepackage{amsmath}
\usepackage{amssymb}
\usepackage{graphicx}
\usepackage{wasysym}
\usepackage[unicode=true]
 {hyperref}

\makeatletter
\renewcommand{\fnum@figure}{FIG.~\thefigure}

\usepackage{algorithm,algpseudocode}
\usepackage{color,xcolor}
\usepackage[caption=false]{subfig}
\definecolor{darkRed}{RGB}{220,0,0}
\definecolor{darkGreen}{RGB}{0,130,0}
\definecolor{darkBlue}{RGB}{0,0,220}
\usepackage{hyperref}
\hypersetup{
      colorlinks=true,
      citecolor=darkGreen,
      linkcolor=darkGreen,
      urlcolor=darkGreen}

\makeatother

\begin{document}

\title{Scaling laws for non-Hermitian skin effect with long-range couplings}

\author{Yi-Cheng Wang}
\affiliation{Department of Physics, National Taiwan University, Taipei 10617, Taiwan}
\affiliation{Institute of Atomic and Molecular Sciences, Academia Sinica, Taipei 10617, Taiwan}
\author{H. H. Jen}
\affiliation{Institute of Atomic and Molecular Sciences, Academia Sinica, Taipei 10617, Taiwan}
\affiliation{Physics Division, National Center for Theoretical Sciences, Taipei 10617, Taiwan}
\author{Jhih-Shih You}
\email{jhihshihyou@ntnu.edu.tw}
\affiliation{Department of Physics, National Taiwan Normal University, Taipei 11677, Taiwan}

\date{\today}

\begin{abstract}
Recent years have witnessed a surge of research on  the non-Hermitian skin effect~(NHSE) in one-dimensional lattices with finite-range couplings. 
In this work, we show that the long-range couplings that decay as $1/l^{\alpha}$ at distance $l$ can fundamentally modify the behavior of NHSE and the scaling of quantum entanglement in the presence of nonreciprocity.
At $\alpha=0$, the nonlocality of couplings gives rise to the scale-free skin modes, whose localization length is proportional to the system size.
Increasing the exponent $\alpha$ drives a complex-to-real spectral transition and a crossover from a scale-free to constant localization length.
Furthermore, the scaling of nonequilibrium steady-state entanglement entropy exhibits a subextensive law due to the nonlocality and the complex spectrum, in contrast to an area law arising from NHSE.
Our results provide a theoretical understanding on the interplay between long-range couplings and non-Hermiticity.
\end{abstract}

\maketitle

{\textit{Introduction.}}---
Nonlocality originating in long-range systems can induce unconventional phases in ground states and quantum dynamics. 
The consequences of tailoring long-range couplings have attracted a variety of research interests, including quantum Hall effect in topological Haldane model~\cite{Haldane1988,Jotzu2014,Yuzhou2021}, localized edge modes in Kitaev's chain~\cite{Vodola2015,Viyuela2016,Patrick2017,Jager2020}, and roton-like dispersion in metamaterials~\cite{Chen2021,Martinez2021}.
The long-range couplings naturally arise in diverse physical platforms~\cite{Defenu2021}, such as trapped ions, Rydberg atoms, and neutral atoms in cavities. These platforms are in general non-Hermitian due to the intrinsic loss process~\cite{Ashida2020}.

Non-Hermiticity can be also induced by the nonreciprocal couplings, which result in an intriguing phenomenon that has no Hermitian counterpart---non-Hermitian skin effect~(NHSE)~\cite{Yao2018,Martinez2018,Kunst2018,Yokomizo2019,Lee2019,Longhi2019,Borgnia2020,Okuma2020,Zhang2020,Kawabata2020,Scheibner2020,Yi2020,Li2020,Zhang2022,Wang2022,Helbig2020,Hofmann2020,Ghatak2020,Weidemann2020,Xiao2020,Xiao2021,Zhang2021,Wang2022,Lin2022NC,Liang2022,Lin2022}.
This means that extensive eigenstates~(skin modes) are exponentially localized at the open boundaries, in stark contrast to the extended Bloch waves under periodic boundary condition.
Intuitively, this is because the nonreciprocal coupling results in the asymmetry in the eigenstate profile, while the periodicity, which provides a tunneling channel between the open ends, smears this asymmetry.
Therefore, an additional long-range coupling can qualitatively alter the localization behavior of NHSE. Besides, understanding its role on the quantum entanglement in the presence of nonreciprocity remains an open question.

In this Letter, we consider a nonlocal model with nonreciprocal and power-law $1/l^{\alpha}$ couplings. 
Once the exponent $\alpha$ is zero, we analytically demonstrate that the localization lengths of extensive skin modes become proportional to the system size for arbitrary system size, which reflects the scale-free localization.
By tailoring the exponent $\alpha$, we find a crossover from a constant to size-dependent localization length as the system size increases, and identify the critical length which indicates the real-to-complex transition in the eigenenergy spectrum.
We further present a simplified model that captures the essence of the real-to-complex spectral transition, which allows us to determine its critical length and the scaling behavior of localization length.
Finally, we demonstrate that nonlocality alters the scaling of nonequilibrium steady-state entanglement entropy, where the area law arising from NHSE becomes the subextensive law as system size increases.

\begin{figure}[!t]
\centering{}
\includegraphics{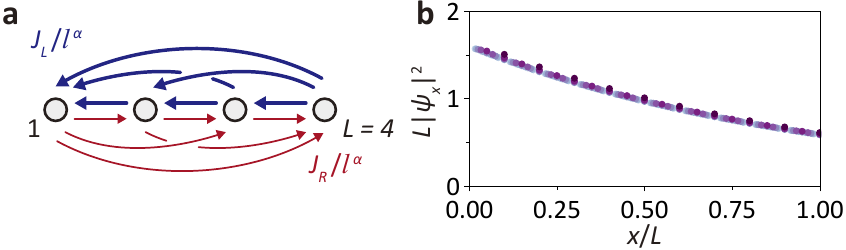}
\caption{\label{fig:Schematics}
(a)~Schematic of 1D lattice consisting of $L=4$ sites with nonreciprocal power-law decaying long-range couplings.
(b)~Rescaled probabilities of the nonlocal model at $\alpha=0,$ $J_L=e^{0.25}$, $J_R=e^{-0.25}$, and $L=10,20,30,40,50,60$.
Coalesence of rescaled probabilities shows the exact scale-free localization at arbitrary system sizes.
}
\end{figure}

{\textit{Nonlocal model.}}---
We start by considering a nonlocal non-Hermitian Hamiltonian under open boundary conditions~(OBC), $\hat{H}_{L,\alpha} =\hat{H}_{L}^\text{HN}+\hat{H}_{L,\alpha}^\text{NL}$~(FIG.~\ref{fig:Schematics}a), where
\begin{align}
\hat{H}_{L}^\text{HN}=\sum_{j=1}^{L-1}\big(J_L\hat{c}_{j}^{\dagger}\hat{c}_{j+1}+J_R\hat{c}_{j+1}^{\dagger}\hat{c}_{j}\big)
\end{align}
corresponds to the Hatano-Nelson model~\cite{Hatano1996,Hatano1997,Hatano1998} and only involves nearest-neighbor nonreciprocal couplings, and
\begin{align}
\hat{H}_{L,\alpha}^\text{NL}=\sum_{j=1}^{L-1}\sum_{l=2}^{L-j}\Big(\frac{J_L}{l^\alpha} \hat{c}_{j}^{\dagger}\hat{c}_{j+l}+\frac{J_R}{l^\alpha} \hat{c}_{j+l}^{\dagger}\hat{c}_{j} \Big)
\end{align}
contains long-range couplings.
Here $\hat{c}_j^\dagger$ ($\hat{c}_j$) is a fermionic creation~(annihilation) operator, $L$ is the number of sites and the power-law decaying exponent $\alpha$ quantifies the strength of nonlocality.

We begin with NHSE at $\alpha = \infty$, where $\hat{H}_{L,\infty}$ reduces to the Hatano-Nelson model $\hat{H}_{L}^\text{HN}$.
The model has single-particle OBC eigenstates $\left | \psi \right \rangle = \sum_{j=1}^{L} \psi_{j} \hat{c}_j^\dagger \left | 0 \right \rangle,$ which exhibit exponentially decaying probability $|\psi_j| \sim e^{-j/\xi_{L,\infty}}$ with size-independent localization length $\xi_{L,\infty}=2(\log{|J_L|/|J_R|})^{-1}$.
In the limit $\alpha \rightarrow 0$~(infinite-range), on the other hand, we are able to derive analytical expressions for the single-particle OBC eigenstate $\left | \psi \right \rangle$ with eigenenergy $E_\text{OBC}=\langle\psi|\hat{H}_{L,\alpha}|\psi\rangle$~\cite{SupplementaryMaterial}.
Here $E_\text{OBC}$ and $\psi_j$ are characterized by an integer $m\in[1,L]$ as
\begin{equation}
\frac{E_\text{OBC}^{(m)}+J_R}{E_\text{OBC}^{(m)}+J_L}=\Big(\frac{J_R}{J_L}\Big)^{1/L}e^{i2\pi\frac{m}{L}}\text{; }\psi_j^{(m)}\propto\Big(\frac{J_R}{J_L}\Big)^{j/L}e^{i2\pi\frac{m}{L}j},\label{eq:OBCwavefunction}
\end{equation}
which exhibits an exponential profile $|\psi_j| \sim e^{-j/\xi_{L,0}}$~(FIG.~\ref{fig:Schematics}b).
In this case, all eigenstates have identical localization length
\begin{equation}
\xi_{L,0}=L\Big(\log{\frac{|J_L|}{|J_R|}}\Big)^{-1}\label{eq:ExactSFL},
\end{equation}
which is proportional to the system size $L$ for arbitrary $L$~(FIG.~\ref{fig:Schematics}b).
Our result signifies the hallmark of the scale-free localization at infinite-range couplings.
Hence, the above discussion implies that the localization length at finite $\alpha>0$ can have nontrivial size-dependence due to the competition between the nearest-neighbor and long-range couplings.

\begin{figure}[!t]
\centering{}
\includegraphics{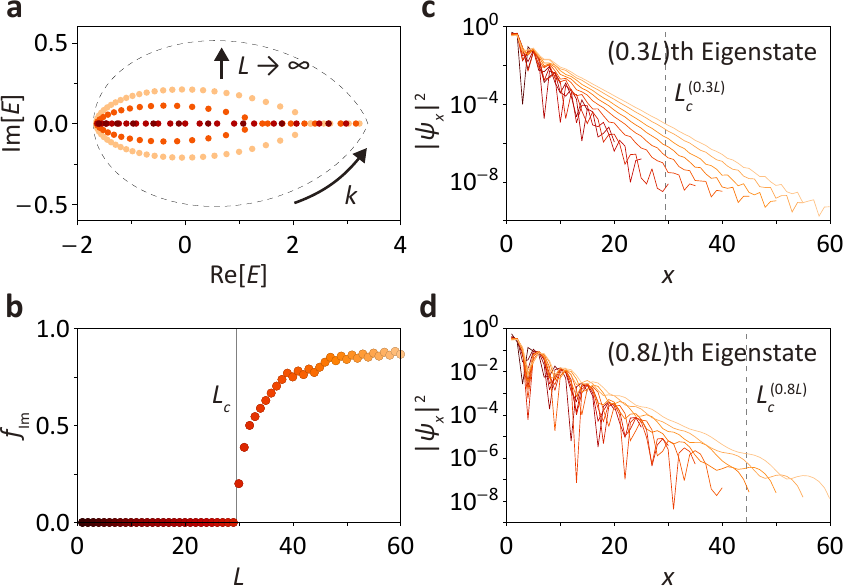}
\caption{\label{fig:SFL}
(a)~Eigenenergies of the nonlocal model at $\alpha=2$, $J_L=e^{0.25}$, $J_R=e^{-0.25}$, and $L=10,20,40,60$.
Dark~(light) color corresponds to small~(large) system size $L$, and OBC spectrum approaches that of an infinite open chain~(dashed) as $L$ increases.
(b)~The fraction of eigenstates with complex eigenenergies shows that $E_\text{OBC}$ undergoes a real to complex transition at the critical length $L_c$~(gray solid line).
(c,d)~Probabilities $|\psi_x|^2$ at~(c) the $(0.3)L$th mode and~(d) $(0.8)L$th mode~(sorted in the ascending order of $\text{Re}[E]$).
}
\end{figure}

{\textit{Effect of exponent $\alpha$.}}---
We now turn to $\alpha>0$ case as shown in FIG.~\ref{fig:SFL}.
To indicate the NHSE here, we first compute the bulk spectrum of a 1D lattice that extends infinitely, such that there is a discrete translational symmetry and the eigenstate can be described by a Bloch wave with momentum $k$.
As a result, the bulk spectrum is given by $E(k)=J_L \text{Li}_\alpha(e^{ik})+J_R\text{Li}_\alpha(e^{-ik})$, where $\text{Li}_\alpha(x)=\sum_{n=1}^\infty x^n/n^\alpha$ is the polylogarithm function.
In FIG.~\ref{fig:SFL}a, the direction of increasing $k$ corresponds to the negative sign of the winding number at $|J_L|>|J_R|$, indicating that there are skin modes localized at the left boundary under OBC~(FIG.~\ref{fig:SFL}c,d)~\cite{Gong2018,Okuma2020}.
Remarkably, we find that by tracking the eigenstates with the same eigenstate index at different system sizes, where eigenstates are sorted in the ascending order of the real parts of eigenenergies, skin modes exhibit the crossover from size-independent to size-dependent localization length.

Another feature we can find in FIG.~\ref{fig:SFL}a is the real-to-complex spectral transition~(FIG.~\ref{fig:SFL}b) in the OBC spectrum as system size increases.
The symmetric spectrum with respect to real energy axis arises since $J_L$ and $J_R$ are real and $\hat{H}_{L,\alpha}$ is pseudo-Hermitian~\cite{Mostafazadeh2002,SupplementaryMaterial}.
As a result, OBC eigenvalues of $H_{L,\alpha}$ are either real or come in complex-conjugated pairs.
In addition, we find that the eigenenergy $E_\text{OBC}^{(m)}$ of $m$th skin mode is real for $L< L_c^{(m)}$ and becomes complex for $L>L_c^{(m)}$. In particular, this critical length $L_c^{(m)}$~(gray dashed lines) also indicates the crossover in the size-dependence of the localization length of $m$th skin modes  in FIG.~\ref{fig:SFL}c,d.
Therefore, we conjecture that there is a correspondence between the real~(complex) OBC eigenenergy and the size-independent~(dependent) localization length.

\begin{figure}[!t]
\centering{}
\includegraphics{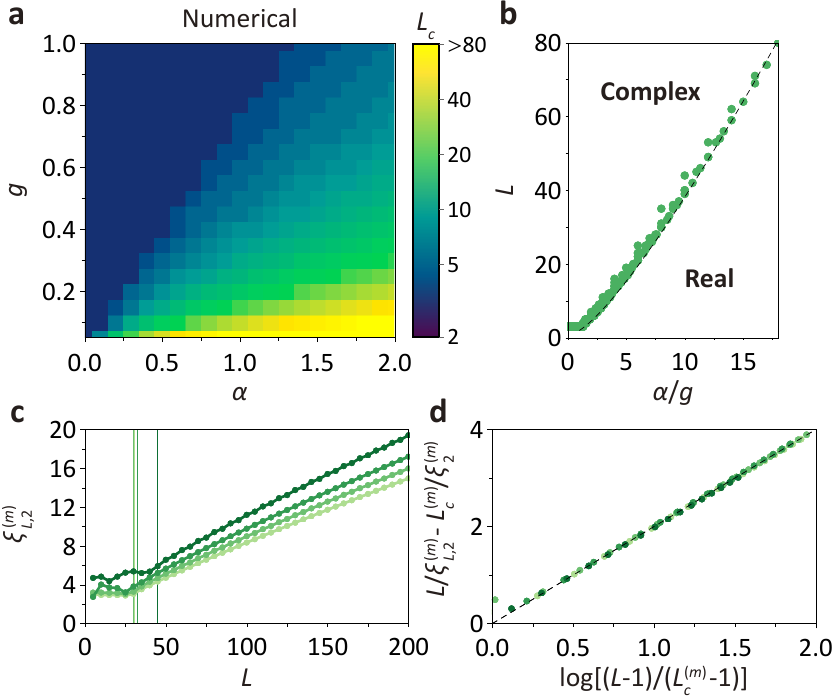}
\caption{\label{fig:CriticalLength}
(a)~Critical length $L_c$ that indicates the real-to-complex spectral transition calculated numerically.
(b)~Critical length as function of $\alpha/g$ obtained numerically~(dots) and according to Eq.~(\ref{eq:CriticalLength})~(dashed line).
(c)~The localization lengths obtained by the fitting function $|\psi_x|\sim e^{-x/\xi_{L,\alpha}^{(m)}}$ at the first, $(0.2L)$th, $(0.4L)$th, $(0.6L)$th, and $(0.8L)$th eigenstates of $\hat{H}_{L,\alpha}$~(light to dark gradient).
The vertical lines represent the corresponding critical lengths $L_c^{(m)}$. (d)~Finite size scaling collapse of $\xi_{L,\alpha}^{(m)}$ at $L> L_c^{(m)}$ according to Eq.~(\ref{eq:LocalizationLength}), where the dashed black line indicates Eq.~(\ref{eq:LocalizationLength}) at $\alpha=2$.
The plots (c) and (d) are obtained with the same parameters in FIG.~\ref{fig:SFL}.
}
\end{figure}

{\textit{Critical length and localization length.}}---
Since $\hat{H}_{L}^\text{HN}$ exhibits the real spectrum and size-independent localization length, we suggest that $H_{L,\alpha}=\hat{H}_{L}^\text{HN}+\hat{H}_{L,\alpha}^\text{NL}$ at small system size is dominated by $\hat{H}_{L}^\text{HN}$.
This means that we can treat $\hat{H}_{L,\alpha}^\text{NL}$ as the correction term to study the occurrence of real-to-complex spectral transition~(FIG.~\ref{fig:CriticalLength}a). Using the imaginary gauge transformation~(IGT), $\hat{c}_j\to e^{-gj}\hat{c}_j$ and $\hat{c}_j^\dagger\to e^{gj}\hat{c}_j^\dagger$ with $g=\ln\sqrt{|J_L|/|J_R|}$, we can thereby map $\hat{H}_{L}^\text{HN}$ to a Hermitian Hamiltonian $J\sum_{j=1}^{L-1}(\hat{c}_{j}^{\dagger}\hat{c}_{j+1}+\hat{c}_{j+1}^{\dagger}\hat{c}_{j})$ and transform the long-range couplings $J_{L(R)}l^{-\alpha}\to J e^{\mp(l-1)g}l^{-\alpha}$.
Here we consider positive $g$ corresponding to $J_L>J_R$ without loss of generality.
Next, we keep the largest rightward coupling $J e^{(L-2)g}(L-1)^{-\alpha}$ and discard the rest coupling terms in $\hat{H}_{L,\alpha}^\text{NL}.$ As a result, $H_{L,\alpha}$ is reduced to a simplified Hamiltonian,
\begin{equation}
\hat{H}_{L,\alpha}^\prime = J\sum_{j=1}^{L-1}(\hat{c}_{j}^{\dagger}\hat{c}_{j+1}+\hat{c}_{j+1}^{\dagger}\hat{c}_{j})+J\mu_L\hat{c}_L^\dagger\hat{c}_1\label{eq:ApproxModel}
\end{equation}
with $\mu_L=e^{(L-2)g}(L-1)^{-\alpha}$ that couples two ends unidirectionally.

The single-particle OBC spectrum of $\hat{H}_{L,\alpha}^\prime$ is given by~\cite{Guo2021,SupplementaryMaterial} $E_\text{OBC}^\prime=2J\cos{\theta}$ with $\sin{[(L+1)\theta]}=\mu_L\sin{\theta}$, and the single-particle wave function is $\psi_j^\prime\propto\sin(j\theta)$.
It is easy to see that the real~(complex) $E_\text{OBC}^\prime$ corresponds to the real~(complex) $\theta$.
At $\mu_L$ smaller~(larger) than one, there are exactly~(less than) $L$ real solutions of $\theta$ within the interval $[0,\pi)$, meaning that OBC eigenspectrum turns complex at $\mu_L=1$. Since the eigenspectrum is invariant under the IGT, the real-to-complex spectral transition in $\hat{H}_{L,\alpha}$ could be indicated by $L_c$ satisfying $\mu_{L_c}=1$, i.e.,
\begin{equation}
e^{(L_c-2)g}=(L_c-1)^\alpha.\label{eq:CriticalLength}
\end{equation}
This criterion implies that when $\mu_L>1$~($L>L_c$),  Eq.~(\ref{eq:ApproxModel}) is dominated by long-range couplings. Therefore the spectral transition can be regarded as the crossover in the dominant coupling term from the nearest-neighbor one to the longest one.
In FIG.~\ref{fig:CriticalLength}b, $L_c$ as a function of $\alpha/g$, obtained numerically from $\hat{H}_{L,\alpha}$ as shown in FIG.~\ref{fig:SFL}b,  is in good agreement with the prediction of Eq.~(\ref{eq:CriticalLength}) shown in dashed line.
The positive correlation between $L_c$ and $\alpha/g$ can be understood from two facts: (i) strong nonlocality~(small $\alpha$) makes long-range couplings dominate the system at a relatively small $L$ and (ii) weak non-Hermiticity~(small $g$) features the real spectrum, such that the spectrum turns complex at a larger $L$.

Our simplified model also allows us to study the size-dependence of the localization length analytically.
We first note that a real $\theta^{(m)}$ corresponds to an extended mode of $\hat{H}_{L,\alpha}^\prime$, showing that the $m$th eigenstate $|\psi^{(m)}\rangle$ of $\hat{H}_{L,\alpha}$ at $L<L_c^{(m)}$ has the size-independent localization length $\xi_{\alpha}^{(m)}$.
Here we emphasize that  each eigenstate  has different $\xi_{\alpha}^{(m)}$ and $L_c^{(m)}$~(see Fig.~\ref{fig:SFL}c,d).
Following the argument which associates $L_c$ with $1/g,$ one can derive the expression of $L_c^{(m)}$ by replacing $g$ in the IGT with $1/\xi_{\alpha}^{(m)}$~\cite{SupplementaryMaterial}.
Under this IGT and at $L>L_c^{(m)},$  the rightward couplings with range larger than $L_c^{(m)}$ are not negligible since their magnitudes are larger than that of
nearest-neighbor couplings.
This relates the wavefunction at each site $\psi^{\prime(m)}_j$ to many other $\psi^{\prime(m)}_l$ due to the nonlocality of long-range couplings.
As a result, the localization length of $|\psi^{(m)}\rangle$ at $L>L_c^{(m)}$ can be approximated by~\cite{SupplementaryMaterial}
\begin{equation}
\xi_{L,\alpha}^{(m)}\sim L\bigg(\frac{L_c^{(m)}}{\xi_{\alpha}^{(m)}}+\alpha\log{\frac{L-1}{L_c^{(m)}-1}}\bigg)^{-1}.\label{eq:LocalizationLength}
\end{equation}
The overall factor $L$ shows a feature of scale-free localization~(FIG.~\ref{fig:CriticalLength}c), while the rest part has an additional logarithmic size-dependence due to the finite $\alpha$.
In FIG.~\ref{fig:CriticalLength}d, we follow Eq.~(\ref{eq:LocalizationLength}) to numerically fit the constant $L_c^{(m)}/\xi_{\alpha}^{(m)}$ from $L/\xi_{L,\alpha}^{(m)}$ with $L\in[L_c^{(m)},200]$ and $\alpha=2$, where $L/\xi_{L,\alpha}^{(m)}$ at $L>L_c^{(m)}$ exhibits a logarithmic size-dependence.
Remarkably, the collapse of numerical results coincides with our prediction in Eq.~(\ref{eq:LocalizationLength}), indicating that the prefactor of logarithmic size-dependence is $\alpha$.

This simplified model is useful to locate the critical length of the real-to-complex spectral transition and to determine the scaling laws of the localization length.
The actual localization length and the OBC spectrum rely on details of all couplings.
Nevertheless, the analysis based on the simplified model agrees well with the numerical calculation~(FIG.~\ref{fig:CriticalLength}) and therefore allows us to conceptually understand the interplay of the non-Hermiticity and the long-range couplings, which is challenging via numerical investigations.

{\textit{Entanglement scalings.}}---
Since the non-Hermiticity gives rise to the nonunitary process that crucially affects the dynamics of a quantum system, it is intriguing to study the role of NHSE in the entanglement dynamics.
Recently, it was reported that NHSE arising from real nonreciprocal nearest-neighbor couplings can lead to the area law of entanglement entropy~\cite{Kawabata2022} in free-fermionic systems under continuous measurement with no-jump condition~\cite{Daley2014}.
Once we incorporate long-range couplings that allow quantum information to propagate between particles with arbitrary separation, 
 the entanglement entropy could be enhanced and exhibit (sub)extensive scaling behavior.
To show this, we consider free fermions on a half-filled open chain and initialize the system in a charge-density-wave state $\prod_{j=1}^{L/2}\hat{c}_{2j}^\dagger|0\rangle.$ Therefore, the time evolution is encoded in the correlation matrix due to the initial Gaussian state and the quadratic form of the model~\cite{Cao2019,Alberton2021,SupplementaryMaterial}.
In this case, the von Neumann entanglement entropy of a wavefunction $|\psi\rangle$ on an $L$-site chain after a bipartition into $A$~($l$ sites) and $B$~($L-l$ sites), $S_{\text{vN},\alpha}(l,L)=-\text{Tr}(\rho_A\log\rho_A)$, can be obtained from the reduced density matrix $\rho_A=\text{Tr}_B|\psi\rangle\langle\psi|$.

\begin{figure}[!t]
\centering{}
\includegraphics{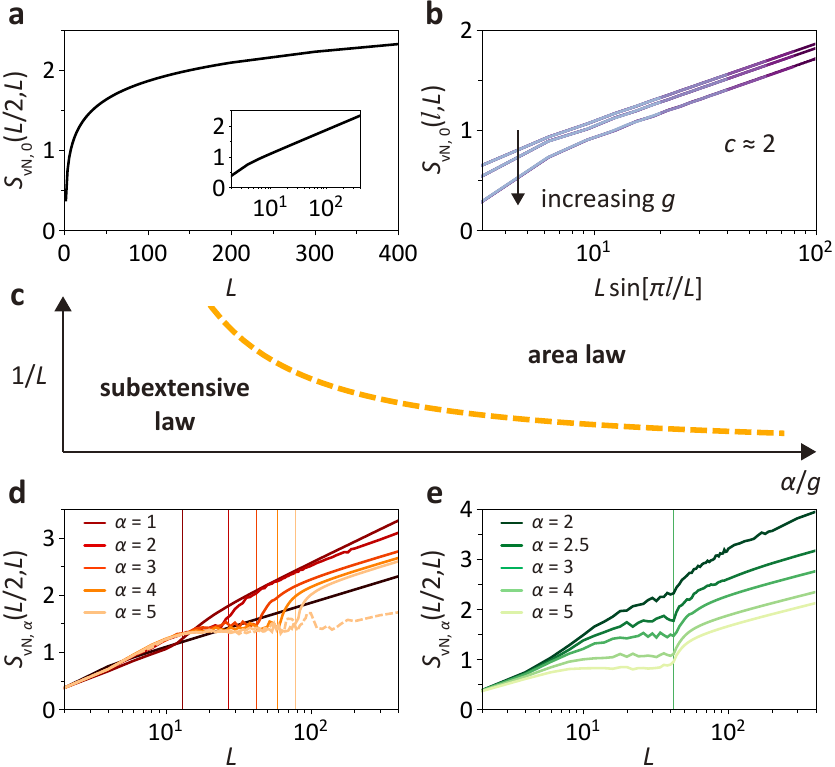}
\caption{\label{fig:EEScaling}
(a)~Subextensive growth of steady-state entropy at half-filling, $g=0.3$, and $\alpha=0$.
Inset: logarithmic size dependence of $S_{\text{vN},0}(L/2,L)$.
(b)~The entropy $S_{\text{vN},0}(l,L)$ as a function of the subsystem size $l$ at $g=0.3$~(top), $0.6$, and $1.2$~(bottom) and $L=20$~(light), $40$, $60$, $80$, and $100$~(dark).
The CFT behaviors can be found at a wide range of $l$.
(c)~Schematic of the size-dependence of entanglement entropy with dashed line indicating the finite size crossover.
(d)~At a fixed $g=0.3$, the half-chain entropies show the subextensive growth $~\log(L)$ and area law $~s_0(g)$ at $\alpha=0$~(black) and $\alpha=\infty$~(dashed), respectively.
At a finite $\alpha$, a crossover from an area law~($L\gtrsim 10$) to an subextensive growth is observed as system size increases.
(e)~With the fixed ratio $\alpha/g=10$ but varying $\alpha$, the abrupt changes that indicates the crossover in the entanglement scaling occur roughly at the same system size.
In (d,e), the crossover is indicated by the vertical lines at which a half of the single-particle spectrum turns complex.
}
\end{figure}

In FIG.~\ref{fig:EEScaling}a, the steady-state entanglement entropy shows a subextensive growth $S_{\text{vN},0}(L/2,L)\sim\log L$, which is reminiscent of a quantum critical system in the $(1+1)$-dimensional conformal field theory~(CFT)~\cite{Calabrese2004}.
To describe this CFT behavior, we calculate the subsystem size dependence of $S_{\text{vN},0}(l,L)$ in FIG.~\ref{fig:EEScaling}b.
The coalesence of $S_{\text{vN},0}(l,L)$ at several $L$ reveals a clear scaling rule with open boundaries
\begin{equation}
S_{\text{vN},0}(l,L)=\frac{c}{6}\log\bigg[\frac{2L}{\pi}\sin\Big(\frac{\pi l}{L}\Big)\bigg]+s_0(g),
\end{equation}
where $c$ is the effective central charge of the corresponding CFT and $s_0$ is a non-universal term.
From FIG.~\ref{fig:EEScaling}b, we numerically obtain a universal $c\approx 2$ for several $g$, and the $g$ dependence lies in $s_0$.
This suggests that the CFT behavior occurs irrespective of NHSE and NHSE results in area law entanglement scaling $s_0(g)$.
Accordingly, the CFT behavior can also be found at $|J_R|=|J_L|$~(no NHSE) as long as $\hat{H}_{L,0}$ is non-Hermitian, where single-particle OBC eigenstates are Bloch waves~(Eq.~(\ref{eq:OBCwavefunction})).
Due to the complex spectrum, the long-time steady state is given by the many-body eigenstate with the largest imaginary part of eigenenergy and therefore resembles the ground state of a tight-binding fermionic model under periodic boundary condition~\cite{SupplementaryMaterial}.
Since the tight-binding fermionic model is described by a CFT with $c_\text{TB}=1$ and its logarithmic entanglement scaling under periodic boundary condition has a prefactor $c_\text{TB}/3$, the central charge for $\hat{H}_{L,0}$ under OBC is given by $c/6=c_\text{TB}/3$ with $c=2$.

These results show that nonlocality along with complex spectrum can give rise to a subextensive growth of the entropy, in stark contrast with the area law in the presense of NHSE with real spectrum at $\alpha=\infty$.
From the competition between nearest-neighbor couplings and long-range couplings, we anticipate the nontrivial finite size effect at $\alpha>0$ as depicted in FIG.~\ref{fig:EEScaling}c.
To reveal this, we first fix the nonreciprocal coupling and study the effect of $\alpha$ in FIG.~\ref{fig:EEScaling}d.
It is evident that abrupt changes in the size dependence of $S_{\text{vN},\alpha}(L/2,L)$ occur at larger system size as $\alpha$ increases.
Before this abrupt change, the entropy follows an area law if $L$ is large enough, otherwise there is no clear entanglement scaling.
This is because the system here is dominated by nearest-neighbor couplings, such that it behaves more like the system at $\alpha=\infty$.
After the abrupt change, the entropy features a logarithmic scaling at large system size, where the system is dominated by long-range couplings.
Accordingly, it is expected that this abrupt change in the steady-state entanglement entropy relates to the competition of nearest-neighbor couplings and long-range couplings.
In FIG.~\ref{fig:EEScaling}d,e, we numerically determine the system size~(vertical line) at which a half of the single-particle spectrum turns complex and find the vertical line indeed indicates the crossover from an area law to a logarithmic scaling.
We note that this criterion follows a similar behavior as Eq.~(\ref{eq:CriticalLength}), which allows us to draw a schematic of entanglement scaling solely by $\alpha/g$ in FIG.~\ref{fig:EEScaling}c.  The $\alpha/g$ dependence is also shown in FIG.~\ref{fig:EEScaling}e.
In conclusion, the finite size crossover in entanglement scaling closely relates to the real-to-complex spectral transition, and they both arise from the competition between nearest-neighbor couplings and long-range couplings.

{\textit{Conclusion.}}---We have revealed the roles of long-range couplings in a non-Hermitian system.
The competition between nearest-neighbor couplings and long-range couplings gives rise to the complex-to-real spectral transition and a crossover from a scale-free to constant localization length of skin modes.
Moreover, we find the logarithmic scaling of steady-state entanglement entropy when the system is dominated by long-range couplings, where a CFT behavior is found at infinite-range couplings.
Our work paves the way towards future studies on the non-Hermitian system with long-range couplings, including the role of boundary conditions, interactions, and the notion of measurement-induced phase transition~\cite{Minato2022,Block2022,Muller2022}.

We thank Po-Yao Chang and Chien-Yu Chou for valuable discussions.
Y.-C.W. and H.H.J. acknowledge support from the National Science and Technology Council~(NSTC), Taiwan, under the Grant No. MOST-109-2112-M-001-035-MY3 and No. MOST-111-2119-M-001-002.
Y.-C.W. and J.-S.Y. are supported by the National Science and Technology Council, Taiwan~(Grant No. MOST-110-2112-M-003-008-MY3).
H.H.J. and J.-S.Y. are also grateful for support from National Center for Theoretical Sciences in Taiwan.

\end{document}